\documentstyle[prb,aps,twocolumn]{revtex}
\begin{document}
\draft

\title{Spin-$\protect\bbox {S}$
bilayer Heisenberg models: Mean-field arguments and
numerical calculations}

\author{Martin P. Gelfand}
\address{Department of Physics, Colorado State University, 
Fort Collins, Colorado 80523}

\author{Zheng Weihong, C.J. Hamer and J. Oitmaa}
\address{School of Physics,                                              
The University of New South Wales,                                   
Sydney, NSW 2052, Australia}

\maketitle

\begin{abstract}
Spin-$S$ bilayer Heisenberg models (nearest-neighbor square
lattice antiferromagnets in each layer, with
antiferromagnetic interlayer couplings) are treated using
dimer mean-field theory for general $S$ and high-order
expansions about the dimer limit for $S=1,\ 3/2,\ldots,4$.
We suggest that the transition between the dimer phase
at weak intraplane coupling and the N\'eel phase at 
strong intraplane coupling is continuous for all $S$,
contrary to a recent suggestion based on Schwinger boson
mean-field theory.  We also present results for $S=1$
layers based on expansions about the Ising limit: 
In every respect the $S=1$ bilayers appear to behave
like $S=1/2$ bilayers, further supporting our picture
for the nature of the order-disorder phase transition.
\end{abstract}

\pacs{PACS numbers:  }

\narrowtext

\section{Introduction and Mean-field Arguments}

The bilayer Heisenberg antiferromagnet,
\begin{equation}
H=J_1 \sum_{a=1,2}\sum_{\langle i,j\rangle}
{\bf S}_{a,i}\cdot {\bf S}_{a,j}
+ J_2 \sum_i {\bf S}_{1,i}\cdot {\bf S}_{2,i}\ ,
\label{eq:Ham}
\end{equation}
where the intralayer couplings run over nearest neighbors
on square lattices and the $\bf S$ are $S=1/2$ quantum spins, 
has turned out to be an excellent
testing ground for notions of quantum criticality.
A variety of controlled calculations, such as strong-coupling
expansions at zero temperature (starting with
the work of Hida\cite{Hida1} and extended
by two of the present authors\cite{Gelfand1,Weihong1}), finite-temperature
quantum Monte Carlo calculations\cite{Sandvik1,Sandvik2} and 
high-temperature expansions\cite{htnote} fit extremely well
into the general field-theoretic framework for zero-temperature
order-disorder critical points in unfrustrated two-dimensional quantum 
antiferromagnets.\cite{Chubukov}

The critical point in the $S=1/2$ bilayer Heisenberg model is between a N\'eel
ordered phase when $\lambda\equiv J_1/J_2$ is sufficiently large
(greater than the critical value $\lambda_c=0.394$) and a
magnetically disordered ``dimer'' phase which is the adiabatic continuation
of the $\lambda=0$ ground state
(which is a product of interlayer singlet pair wavefunctions).
The universality class of the phase transition is the same as
for finite-temperature transitions in $d=3$, $O(3)$ symmetric
models.

It seemed most natural to us that for larger values of the
quantum spin $S$ the same scenario would play out:  The value
of $\lambda_c$ would change, but qualitatively nothing
would be different.  It therefore came as a surprise when
Ng, Zhang, and Ma\cite{Ng} described the results of Schwinger boson
mean-field (SBMF) calculations, which indicate that for 
$S>S_c$ the transition between magnetically ordered and disordered
phases is {\it first-order\/}.  The calculation leads to $S_c\approx0.35$,
and hence indicates that for all physical values of $S$ the transition
should be first-order; but it is reasonable to accept the suggestion
of Ng {\it et al.\ }that SBMF theory leads to phase diagram of
$S$ versus $\lambda$ which is qualitatively valid 
even if $S_c$ is underestimated.  
Two more salient points regarding the SBMF results should be noted.
At large $S$, the value of $\lambda$ at the
dimer-N\'eel phase boundary scales as $1/S$;
and at sufficiently large $S$ the ground state on the dimer
side of the boundary is the pure interlayer singlet pair
state ({\it i.e.,} identical to the ground state at $\lambda=0$).

Here we will argue that the results of SBMF theory for phase
transitions in the bilayer Heisenberg model
are qualitatively incorrect, even in the limit of large $S$ where
one might expect such an approximation to be reliable.

Ng {\it et al.\ }offer an elementary argument in support of their proposed
scenario, which it is informative
to reconsider.  They note that the ideal N\'eel state and the interlayer
singlet state become degenerate for $\lambda=\lambda_1=O(1/S)$
(with the interlayer singlet state energetically
favored for $\lambda<\lambda_1$), while linear
spin-wave theory indicates that the magnetization should vanish
at $\lambda=\lambda_2=O(1/S^2)$ (with N\'eel order existing
for $\lambda>\lambda_2$).  At large $S$, it is clear
that  $\lambda_1>\lambda_2$.  Ng {\it et al.\ }suggest that, on decreasing
$\lambda$ from infinity,
the transition that one might suspect should take place at
$\lambda_2$ (presumably a continuous transition from the
N\'eel to dimer phase) is pre-empted by a first-order transition
at $\lambda_1$.  This is certainly consistent with their SBMF calculations.

However, we suggest an alternative interpretation.
We believe that the argument in favor of a first-order transition
at $\lambda_1$ is too simplistic:  Using simple trial states
to represent the ground states in the respective phases is 
dangerous, particularly since it is a comparison of
subleading (in powers of $S$) terms in the variational energies for 
each phase which leads to the estimate for $\lambda_1$.
Furthermore, there is an elementary calculation which strongly
suggests that the dimer phase is unstable with respect to
the development of N\'eel order for $\lambda>\lambda_3=O(1/S^2)$.
Consider a single dimer, subject to a staggered magnetic field
which is produced self-consistently by the staggered magnetization
of the system.  Let $J_2\chi_0(S)$ denote the staggered susceptibility
of a single dimer.  Then, within this ``dimer mean-field approximation''
one has $\lambda_3=1/4\chi_0$.  One can calculate $\chi_0$ directly
from second-order perturbation theory,
based on explicit formulae for 3-$j$ symbols,\cite{Rotenberg}
to find that $\lambda_3=3/16S(S+1)$.  As promised, the domain
of stability of the dimer phase is $O(1/S^2)$, in agreement with
the spin-wave estimate for the stability of the N\'eel phase.

It should be clear that the dimer mean-field approximation
underestimates the stability of the dimer phase, just as
the standard Weiss theory overestimates Curie temperatures.
One could reasonably object to our argument, then, by suggesting
that the corrections to this mean-field theory could be large:
in particular, $\lambda_3$ might be pushed up to a value of order $1/S$.
In the following section, we describe high-order perturbation
expansions about $\lambda=0$ for bilayers
with $S=1,\ 3/2,\ldots,\ 4$.
Combined with the existing results for $S=1/2$, they
strongly suggest that the corrections to dimer mean-field theory
are order unity rather than order $S$.  In addition, for $S=1$
bilayers we present a wide variety of results based on 
expansions about $\lambda=0$ and about Ising models.  They
are all qualitatively similar to what is found for $S=1/2$ bilayers,
providing additional confirmation that the phase diagram is
unchanged with increasing $S$ (in particular, there is no evidence
for any phase besides dimer or N\'eel) and offering accurate
results which could prove useful in interpreting data from
experimental realizations of $S=1$ bilayers.

\section{Series expansions and extrapolations}

\subsection{Triplet Gap and Antiferromagnetic Susceptibility}

Perturbation expansions\cite{pertexp} for the ground-state energy $E_0$,
the triplet excitation 
spectrum, and the antiferromagnetic susceptibility $\chi$ have been
carried out for bilayers with all integer and half-integer
values of $S$ from 1 to 4.
% $O(\lambda^9)$ for $S=1$ bilayers, 
% $O(\lambda^7)$ for $S=2$ and $S=3$ bilayers, 
% and $O(\lambda^6)$ for $S=4$  bilayers.  
The series coefficients for integer $S$ bilayers
for the minimum gap $\Delta$, corresponding to wave vector $(\pi,\pi)$,
and $\chi$
are presented in Table~\ref{table:gaps}.
For $S=1/2$ bilayers such expansions have already been
presented by Zheng\cite{Weihong1} to $O(\lambda^{11})$.

Fig.~\ref{gapapprox} is a set of ``scatter plots'' of estimated
critical $\lambda$ values versus estimated critical exponents
derived from unbiased $D$log-Pad\'e approximants to the gap series
for $S=1$, 2, 3 and 4 bilayers.  An analogous plot for $S=1/2$
is presented in Ref.~\onlinecite{Gelfand1}.
In every case they have very similar character,
with nearly all exponent estimates lying slightly above the
anticipated value of 0.71, 
and with clear correlations between critical 
point and exponent values.  
(Scatter plots for the antiferromagnetic susceptibility have
very similar character; likewise for the gap and susceptibility
of half-integer $S$ bilayers.)
As $S$ increases the series actually become better behaved,
in that the approximants derived from a fixed number of terms
are more tightly clustered (neglecting the outliers) 
and estimated exponents are closer to the expected value.
Biasing the exponent one can obtain more precise
estimates of the critical points $\lambda_c$, as
listed in Table~\ref{table:xc}.
% leads to the values
% 0.3942(8), 0.1393(7), 0.07303(10), 0.0453(1), 0.03098(7),
% 0.02254(5), 0.01716(5),   and 0.01349(5) 
% for $S=1/2$ (already reported in
% Ref.~\onlinecite{Weihong1}), 1, 3/2, 2, 5/2, 3, 7/2
% and 4, respectively.  The ratio
% of these values to those coming from dimer mean-field theory 
% is 1.58, 1.49, 1.46, 1.45, 1.44, 1.44, 1.44 and 1.44.  
The trend is clear:  with increasing
$S$, the critical value of $\lambda$ is approaching a constant
multiple of the dimer mean-field value $3/16S(S+1)$.
If we plot $1/\lambda_c$ as a function of $R\equiv S(S+1)$ (shown in
Fig.~\ref{lambda_c}), we can see that the results can be remarkably
well fitted by a
straight line 
\begin{equation}
1/\lambda_c= c (R - R_c ) \label{fit_lambdac}
\end{equation}
with an intercept at $1/\lambda_c=0$
slightly above $R=0$. The constants $c$ and $R_c$ 
determined by a linear least squares fit are:
\begin{equation}
c=3.72(1),\quad   R_c=0.068(2)~. \label{R_c}
\end{equation}
% so $S_c=0.064(2)$.

% We can be more precise concerning this last point.  
For general $S$ we have found empirically,
based on the ground state energy $E_0$, the minimum triplet gap $\Delta$,
and the antiferromagnetic susceptibility $\chi$ series for the eight values of
$S$ that have been explicitly calculated, that the coefficients
of $\lambda^n$ can be expressed as polynomials of order $n$
in the variable $R$.
%$R\equiv S(S+1)$. 
For example, to third order
\begin{eqnarray}
&&\Delta=1-(8/3)R\lambda+[(8/9)R-(32/27)R^2]\lambda^2 \label{delta}\\
&&+[(56/135)R-(116/135)R^2-(3824/1215)R^3]\lambda^3\ .\nonumber
\end{eqnarray}
Further coefficients are listed in the Appendix.
%\footnote{
%Eq.~(\ref{delta}) is naturally consistent with the linear
%form (\ref{fit_lambdac}), given that the expected
%critical index is independent of $S$; any other terms (such as
%$R^2$ or $1/R$) in Eq.~(\ref{fit_lambdac}) would tend to imply terms of
%order \{$R^m\lambda^n, m>n {\rm ~or~} m<0$\} in the double series
%(\ref{delta}).} 
One can then define $r=\lambda R$ and consider the $R\to\infty$, 
$r\to\rm const$ limit.
The resulting series in $r$ corresponds to the terms in the double-series
(in $R$ and $\lambda$) of the form $R^n\lambda^n$.  Its $D$log-Pad\'e
approximants are as well behaved as the series for the larger values
of $S$ shown above, with unbiased approximants clustering and
exponent estimates just above the anticipated value.  Biasing the
critical point estimates leads to a critical $r$ of 0.2691(7)
(this is consistent with the value of $c$ in Eq.~(\ref{R_c})),
so the ratio of the exact critical point value in the large-$S$ limit
to that coming from the dimer mean-field theory is 1.435(4).

We can also try to use the double-expansions in $R$ and $\lambda$
to analytically continue to unphysically small values of $S$.
An interesting feature of the general-$S$ double-series (both for
the $\Delta$ and $\chi$) is
that all terms of the form $R^0 \lambda^{n\ge1}$ appear to have vanishing
coefficients.  Consequently, for $S=0$ there is no phase transition
to a N\'eel-ordered state.  Presumably there is a critical spin
$S_c$, less than $1/2$, at and below which the dimer state is 
stable for all values of $\lambda$.  
It has not been possible to obtain a reliable estimate of $S_c$
by constructing $D$log-Pad\'e approximants at fixed values of $S$:
for $S\alt 0.35$ the approximants depend strongly on the number
of terms used. However, the extrapolation from
larger $S$, summarized in Eqs.~(\ref{fit_lambdac}) and
(\ref{R_c}), suggests that $S_c=0.064(2)$.

\subsection{Further results for $\protect\bbox{S=1}$ bilayers}

Although we have presented evidence above that the dimer phase is unstable
for $\lambda$ larger than a critical value of $O(1/S^2)$, we did not
firmly establish that the structure of the phase diagram is 
simply dimer phase--critical point--N\'eel phase.  
The fact that the unbiased exponent values are consistent with
a Lorentz-invariant, $d=3$, $O(3)$ universality class is certainly
suggestive.  For the $S=1$ bilayers we have done much more.
In addition to expansions about $\lambda=0$ we have constructed
expansions about Ising models, generating a set of results
analogous to those presented by Zheng\cite{Weihong1} for
$S=1/2$ bilayers.  Taken together, these leave little room for
doubt about the phase diagram.

Because the calculations follow those presented in Ref.~\onlinecite{Weihong1}
so closely we refer the reader to Sec.~III of that paper for
a description of the calculation (note that the parameter $y$ in that paper
corresponds to $1/\lambda$).  Expansions were obtained to $O(x^{9})$
(with $x$ denoting the ratio of transverse to longitudinal exchange
strength)
for the sublattice magnetization $M$, uniform transverse susceptibility
$\chi_\perp$, and triplet excitation spectrum. From these we could
derive the gap to the optical branch of the spin-wave
spectrum at wave vectors $(0,0)$ ($\Delta_{\rm opt}$, the
minimum gap) and $(0,\pi)$ ($\Delta_X$),
spin-wave velocity $v$, and spin-wave stiffness $\rho_s$.
The expansion coefficients will not be presented here but
are available from the authors on request.  Results of
series extrapolations are shown in Figs.~\ref{optical_gap},
\ref{M_chi}, and \ref{c_rho}.  They demonstrate that expansions
about the N\'eel state lead to estimates of the domain of stability of the
N\'eel phase which are completely consistent with the critical point 
found by expanding about $\lambda=0$.
This is nicely confirmed by Fig.~\ref{e0}, which displays the
ground-state energy per site. The values from the dimer expansion
match very smoothly on to those from the Ising expansion around
the critical point, whereas for a first-order transition there would be
a discontinuity in slope at the transition.
All of the results presented in this subsection represent the 
best available theoretical values for
experimentally accessible properties of N\'eel-ordered $S=1$ bilayers,
which should prove useful in the interpretation of experimental
data if any such systems are studied.  (In real compounds single-ion
anisotropy is always present to some degree. The present calculations
could readily be extended to include such terms in the Hamiltonian.)

Finally, it is amusing to note that in the dimer phase for 
$S>1/2$ bilayers, sufficiently close to $\lambda=0$, there
exist spin-2 elementary excitations.
Let us consider $S=1$ bilayers
specifically.  Triplet spectra for various values of
$\lambda$ in the dimer phase are shown in Fig.~\ref{mk_triplet}
while {\it quintuplet\/} spectra are shown in Fig.~\ref{mk_quintuplet}.
(These spectra are obtained by direct sums of the terms in
the series to the maximum order available.)
The latter spectra lose physical significance when the quintuplet
excitations can decay into multiple triplet excitations. 
One might imagine this happens for arbitrarily small values of $\lambda$,
since at $\lambda=0$ the quintuplet gap is three times the
triplet gap.  However, the parity, with respect to interchange of 
layers in the bilayer system, of the quintuplet excitations
is opposite to that of the triplet excitations, and symmetry forbids
decay of the quintuplet excitations into an odd number of triplet
excitations.  Hence stable spin-2 excitations lie between the
2-triplet and 4-triplet continua for sufficiently small $\lambda$.

\section{Discussion}

The order-disorder transition in bilayer
Heisenberg antiferromagnets appears to
be a problem for which Schwinger boson 
mean-field theory leads to qualitatively incorrect results
even in the limit of large $S$.  We have shown by high order
perturbation expansions about the limit of uncoupled interlayer
singlets that there is strong evidence 
for continuous transitions between dimer and N\'eel phases
with critical values of $J_1/J_2$ scaling as $1/S(S+1)$.
There seems to be no evidence in favor of the scenario
described by Ng {\it et al.},\cite{Ng} in which the transition
would become first-order for sufficiently large $S$.
Of course this does not imply that the phase diagram for bilayer
Heisenberg antiferromagnets is completely universal.  If further-neighbor
interactions or higher-order exchange interactions (for example, terms
in the Hamiltonian of the form $({\bf S}_i\cdot {\bf S}_j)^2$)
were allowed then a wide variety of new phases could be stabilized.
However, for the simplest bilayer models it appears that 
simple arguments based on the instability of the dimer phase to N\'eel order
(using dimer mean-field theory) and the instability of the
N\'eel phase to a spin-disordered phase
(using linear spin-wave theory to determine where
the staggered magnetization vanishes), which both suggest a
critical point at $J_1/J_2=O(1/S^2)$, are correct.

\acknowledgments

This work has been supported by NSF Grant No. DMR 94-57928 (MPG) and
the Australian Research Council (ZW, CJH and JO).

\appendix
\section*{}
The double expansions in $\lambda$
and $R\equiv S(S+1)$ for the ground-state energy per site $E_0/N$,
triplet minimum energy gap $\Delta$, and antiferromagnetic
susceptibility $\chi$ for spin-$S$ bilayers are presented in full here.

\hbox{}
%\widetext
\begin{eqnarray}
{E_0\over N J_2} && = {{-R}\over 2} - {{2\,{{\lambda }^2}\,{R^2}}\over 3} - 
   {{{{\lambda }^3}\,{R^2}}\over 3} \nonumber \\
&&  + {{\lambda }^4}\,{R^2}\,\left( -{1\over 5} + {{28\,R}\over {45}} - 
      {{38\,{R^2}}\over {45}} \right)   \nonumber \\
&& +  {{\lambda }^5}\,{R^2}\,\left( -{2\over {15}} + {{74\,R}\over {75}} - 
      {{151\,{R^2}}\over {225}} \right)   \nonumber \\
&& + {{\lambda }^6}\,{R^2}\, {\Big (} -0.095238095 + 1.18141968\,R  \nonumber \\
&&  -   1.93520995\,{R^2} + 3.52146794\,{R^3} - 2.4209868\,{R^4} {\Big )}   \nonumber \\
&& + {{\lambda }^7}\,{R^2}\, {\Big (} -0.071428571 + 1.28412698\,R  \nonumber \\
&& -  4.19416352\,{R^2} + 5.77329638\,{R^3} - 1.84452422\,{R^4} {\Big )}\nonumber \\
&& + O(\lambda^8) \\
\Delta &&=  1 - {{8\,\lambda \,R}\over 3} 
+ {{\lambda }^2}\,R\,\left( {8\over 9} - {32 R \over 27} \right) \nonumber \\
&& + {{\lambda }^3}\,R\,\left( {{56}\over {135}} - {{116\,R}\over {135}} - 
      {{3824\,{R^2}}\over {1215}} \right)    \nonumber \\
&& + {{\lambda }^4}\,R\,\left( {{20}\over {81}} - {{655\,R}\over {189}} 
   +  {{5120\,{R^2}}\over {1701}} - {{48656\,{R^3}}\over {15309}} \right) \nonumber \\
&& +  {{\lambda }^5}\,R\, {\Big (} 0.166019988 - 4.50447076\,R  \nonumber \\
&& + 8.81922354\,{R^2} + 
      5.13123542\,{R^3} - 14.6722347\,{R^4} {\Big )}    \nonumber \\
&& + {{\lambda }^6}\,R\, {\Big (} 0.112022751 - 5.43708747\,R + 
      24.41683462\,{R^2}  \nonumber \\
&& - 23.43145101\,{R^3} + 17.32806918\,{R^4} - 
      21.97846687\,{R^5} {\Big )}   \nonumber \\
&& + {{\lambda }^7}\,R\, {\Big (} 0.091548205 - 6.26174454\,R + 
      45.07198387\,{R^2} \nonumber \\
&& - 73.28433315\,{R^3} + 29.56605425\,{R^4} + 
      59.68596039\,{R^5} \nonumber \\
&& - 106.4935325\,{R^6} {\Big )}  + O(\lambda^8) \\
% &&\Delta=1-(8/3)R\lambda+[(8/9)R-(32/27)R^2]\lambda^2\\
% &&+[(56/135)R-(116/135)R^2-(3824/1215)R^3]\lambda^3+\cdots\ .\nonumber
\chi &&=  {4R\over 3} {\Big [} 1 + {{16\,\lambda \,R}\over 3} + 
        {{{{\lambda }^2}\,R\,\left( -1 + 140\,R \right) }\over 6} \nonumber \\
&& +  {{\lambda }^3}\,R\,\left( -{1\over 8} - {{143\,R}\over {36}} + 
           {{2677\,{R^2}}\over {27}} \right)    \nonumber \\
&& +   {{\lambda }^4}\,R\, {\Big (} -0.10671296 - 1.45847442\,R \nonumber \\
&&  -   33.2428954\,{R^2} + 403.712863\,{R^3} {\Big )}    \nonumber \\
&& +   {{\lambda }^5}\,R\, {\Big (} -0.090168701 - 0.58931360\,R \nonumber \\
&& -  3.49146301\,{R^2} - 221.1724119\,{R^3} + 1622.218477\,{R^4} {\Big )} \nonumber \\
&& + \! {{\lambda }^6}R {\Big (}\! -0.076292486 + 0.12455955 R + 
           2.92215451 {R^2} \nonumber \\
&& + 8.15178055\,{R^3} - 1231.683801\,{R^4} + 6416.371180\,{R^5}  {\Big )} \nonumber \\
&& + O(\lambda^7) {\Big ]} 
\end{eqnarray}

\begin{figure}
\caption{Plots of critical $\lambda$ values versus critical
exponents for $D$log-Pade approximants to the triplet gap series.
All approximants that use all available terms in the series
(circles) and all but the last term (squares) are displayed.
Note that many of the approximants are difficult to distinguish
on these plots, but the ``outliers'' represent single approximants.}
\label{gapapprox}
\end{figure}

\begin{figure}
\caption{Plot of the inverse of critical point ($1/\lambda_c$)
versus $S(S+1)$: the crosses with error bars
are the estimates biased by the critical index, and the dashed line
is a linear fit: $1/\lambda_c=3.72 [S(S+1) - 0.068]$.
The insert enlarges the region near $S=0$.
}
\label{lambda_c}
\end{figure}

\begin{figure}
\caption{Rescaled energy gaps $\Delta_{\rm opt}$ and
$\Delta_{\rm X}$  as a
function of $J_2/(J_1+J_2)$.
The solid curves at large $J_2/(J_1+J_2)$
are extrapolations based on dimer expansions;
the crosses with error bars are the estimates
from Ising expansions; the long dashed lines at small $J_2/(J_1+J_2)$ 
are the results of the linear spin-wave theory;\protect\cite{swexp} and
the open circles connected by a short dashed line are results from
the theory of Millis and Monien\protect\cite{mil96} (with
$\rho_s$, an input to their theory, 
taken from Fig.~\protect\ref{c_rho}).
The vertical dotted line indicates the critical point.}
\label{optical_gap}
\end{figure}

\begin{figure}
\caption{The staggered magnetization $M$ and 
uniform perpendicular susceptibility $\chi_{\perp}$
for $S=1$ bilayers {\it versus\/} $(J_2/J_1)^{1/2}$ as 
estimated by Ising expansions.}
\label{M_chi}
\end{figure}

\begin{figure}
\caption{The  spin-wave velocity  $v$
and  the spin-wave stiffness $\rho_s$ 
{\it versus\/} $(J_2/J_1)^{1/2}$ as estimated by Ising expansions.
At the critical ratio $J_2/J_1=7.18$, the estimate of $v$ from
dimer expansions is also shown.}
\label{c_rho}
\end{figure}

\begin{figure}
\caption{The rescaled ground-state energy per site 
$E_0/(4J_1+J_2)/N$ as a function of $J_2/(J_1+J_2)$.
The crosses with dashed line connecting them
 are the results from the expansion about the
Ising limit, while the solid line gives the results from
the dimer expansion.
}
\label{e0}
\end{figure}

\begin{figure}
\caption{Plot of the  spin-triplet excitation spectrum 
$\Delta(k_x,k_y)$  along high-symmetry
cuts through the Brillouin zone for the system with coupling ratios
$J_1/J_2=0.05$, 0.1, 0.12, 0.1393 (shown in
the figure from  the top to 
the bottom at $(\pi,\pi)$  respectively). 
The lines are the estimates by direct sum to the dimer series,
and the points (circles with error bar 
for the case of $J_1/J_2=0.1393$ only) are the 
estimates of the Pad\'{e} approximants to the dimer series.}
\label{mk_triplet}
\end{figure}

\begin{figure}
\caption{Plot of the quintuplet  excitation spectrum 
$\Delta^q(k_x,k_y)$ along high-symmetry
cuts through the Brillouin zone for the system with coupling ratios
$J_1/J_2=0.05$, 0.1, 0.12, 0.1393.  
The lines are the estimates by direct sum to the dimer series,
and the points (circles with error bar 
for the case of $J_1/J_2=0.1393$ only) are the estimates 
of the Pad\'{e} approximants to the dimer series.}
\label{mk_quintuplet}
\end{figure}

\newpage
\hbox{}
\newpage

\mediumtext
\begin{table}
\squeezetable
\setdec 0.0000000000000
\caption{Series coefficients $c_n$ for dimer expansions of  
the minimum triplet
gap $\Delta=J_2\sum_n c_n \lambda^n$ and the antiferromagnet
susceptibility $\chi=J_2^{-1} \sum_n c_n \lambda^n$
for $S=1$, 2, 3, and 4 bilayers.}
\label{table:gaps}
\begin{tabular}{rrrrr}
\multicolumn{1}{c}{$n$} &\multicolumn{1}{c}{$S=1$}
&\multicolumn{1}{c}{$S=2$}&\multicolumn{1}{c}{$S=3$}&\multicolumn{1}{c}{$S=4$}  \\
\tableline
\multicolumn{1}{c}{}&\multicolumn{4}{c}{Minimum triplet gap} \\
  0 &\dec  1.0000000000   &\dec  1.0000000000      
 &\dec  1.0000000000      &\dec  1.0000000000      \\
  1 &\dec $-$5.3333333333      &\dec $-$1.6000000000$\times 10^{1}$
 &\dec $-$3.2000000000$\times 10^{1}$ &\dec $-$5.3333333333$\times 10^{1}$ \\
  2 &\dec $-$2.9629629630      &\dec $-$3.7333333333$\times 10^{1}$
 &\dec $-$1.6000000000$\times 10^{2}$ &\dec $-$4.5629629630$\times 10^{2}$ \\
  3 &\dec $-$2.7786008230$\times 10^{1}$ &\dec $-$7.0826666667$\times 10^{2}$
 &\dec $-$5.5573333333$\times 10^{3}$ &\dec $-$2.5514008230$\times 10^{4}$ \\
  4 &\dec $-$4.0140832190$\times 10^{1}$ &\dec $-$3.5921481481$\times 10^{3}$
 &\dec $-$6.1199238095$\times 10^{4}$ &\dec $-$4.8582313672$\times 10^{5}$ \\
  5 &\dec $-$3.3454379922$\times 10^{2}$ &\dec $-$1.0569742866$\times 10^{5}$
 &\dec $-$3.5299272466$\times 10^{6}$ &\dec $-$4.6061398131$\times 10^{7}$ \\
  6 &\dec $-$1.0532000998$\times 10^{3}$ &\dec $-$9.1597242219$\times 10^{5}$
 &\dec $-$6.1760036073$\times 10^{7}$ &\dec $-$1.3547279280$\times 10^{9}$ \\
  7 &\dec $-$9.7019942999$\times 10^{3}$ &\dec $-$2.6882225528$\times 10^{7}$
 &\dec $-$3.6317201466$\times 10^{9}$ &                \\
  8 &\dec $-$3.0018452174$\times 10^{4}$ &                       &     
                      &                \\
  9 &\dec $-$3.0081157280$\times 10^{5}$ &                       &   
                        &                \\
\tableline
\multicolumn{1}{c}{}&\multicolumn{4}{c}{Antiferromagnetic susceptibility} \\
  0 &\dec  2.6666666667       &\dec  8.0000000000       &\dec 
 1.6000000000$\times 10^{1}$ &\dec  2.6666666667$\times 10^{1}$ \\
  1 &\dec  2.8444444444$\times 10^{1}$ &\dec  2.5600000000$\times 10^{2}$
 &\dec  1.0240000000$\times 10^{3}$ &\dec  2.8444444444$\times 10^{3}$ \\
  2 &\dec  2.4800000000$\times 10^{2}$ &\dec  6.7120000000$\times 10^{3}$
 &\dec  5.3728000000$\times 10^{4}$ &\dec  2.4880000000$\times 10^{5}$ \\
  3 &\dec  2.0721234568$\times 10^{3}$ &\dec  1.7017800000$\times 10^{5}$
 &\dec  2.7320720000$\times 10^{6}$ &\dec  2.1109167901$\times 10^{7}$ \\
  4 &\dec  1.6499774192$\times 10^{4}$ &\dec  4.1278260778$\times 10^{6}$
 &\dec  1.3301975846$\times 10^{8}$ &\dec  1.7154007839$\times 10^{9}$ \\
  5 &\dec  1.2891136900$\times 10^{5}$ &\dec  9.8615644145$\times 10^{7}$
 &\dec  6.3850802835$\times 10^{9}$ &\dec  1.3748488989$\times 10^{11}$ \\
  6 &\dec  9.9036806734$\times 10^{5}$ &\dec  2.3183667236$\times 10^{9}$
 &\dec  3.0164597503$\times 10^{11}$ &\dec  1.0845538535$\times 10^{13}$ \\
  7 &\dec  7.5397998426$\times 10^{6}$ &\dec  5.4074404771$\times 10^{10}$
 &       &       \\
  8 &\dec  5.6895542718$\times 10^{7}$ &       &      &      \\
\end{tabular}
\end{table}

\mediumtext
\begin{table}
\squeezetable
\setdec 0.00000
\caption{The critical point $\lambda_c$ for $S=1/2$ to $S=4$
estimated by biasing the critical index 
(where the case of $S=1/2$ is already reported in 
Ref.~\protect\onlinecite{Weihong1}),
and its ratio to that predicted by dimer mean-field theory.
The results of a linear fit, 
$1/\lambda_c^{{\rm fit}} = 3.72 [ S (S+1)-0.068]$,
are also listed.
}
\label{table:xc}
\begin{tabular}{ccccccccc}
\multicolumn{1}{c}{$S$} &\multicolumn{1}{c}{$1/2$}
&\multicolumn{1}{c}{$1$}&\multicolumn{1}{c}{$3/2$}&\multicolumn{1}{c}{$2$}
&\multicolumn{1}{c}{$5/2$}&\multicolumn{1}{c}{$3$}&\multicolumn{1}{c}{$7/2$}
&\multicolumn{1}{c}{$4$}\\
\tableline
$\lambda_c$ & 0.3942(8) & 0.1393(7) & 0.07303(10)& 0.0453(1)&
 0.03098(7) & 0.02254(5) & 0.01716(5) & 0.01349(5) \\
$\lambda_c^{{\rm fit}}$ & 0.3942 & 0.1391 & 0.07301 & 0.04532 & 0.03096 &
0.02253 & 0.01714 & 0.01349 \\
$\lambda_c/\lambda_3$ &   1.58 & 1.49 & 1.46 & 1.45 & 1.44 & 1.44 &
 1.44 & 1.44 \\
\end{tabular}
\end{table}

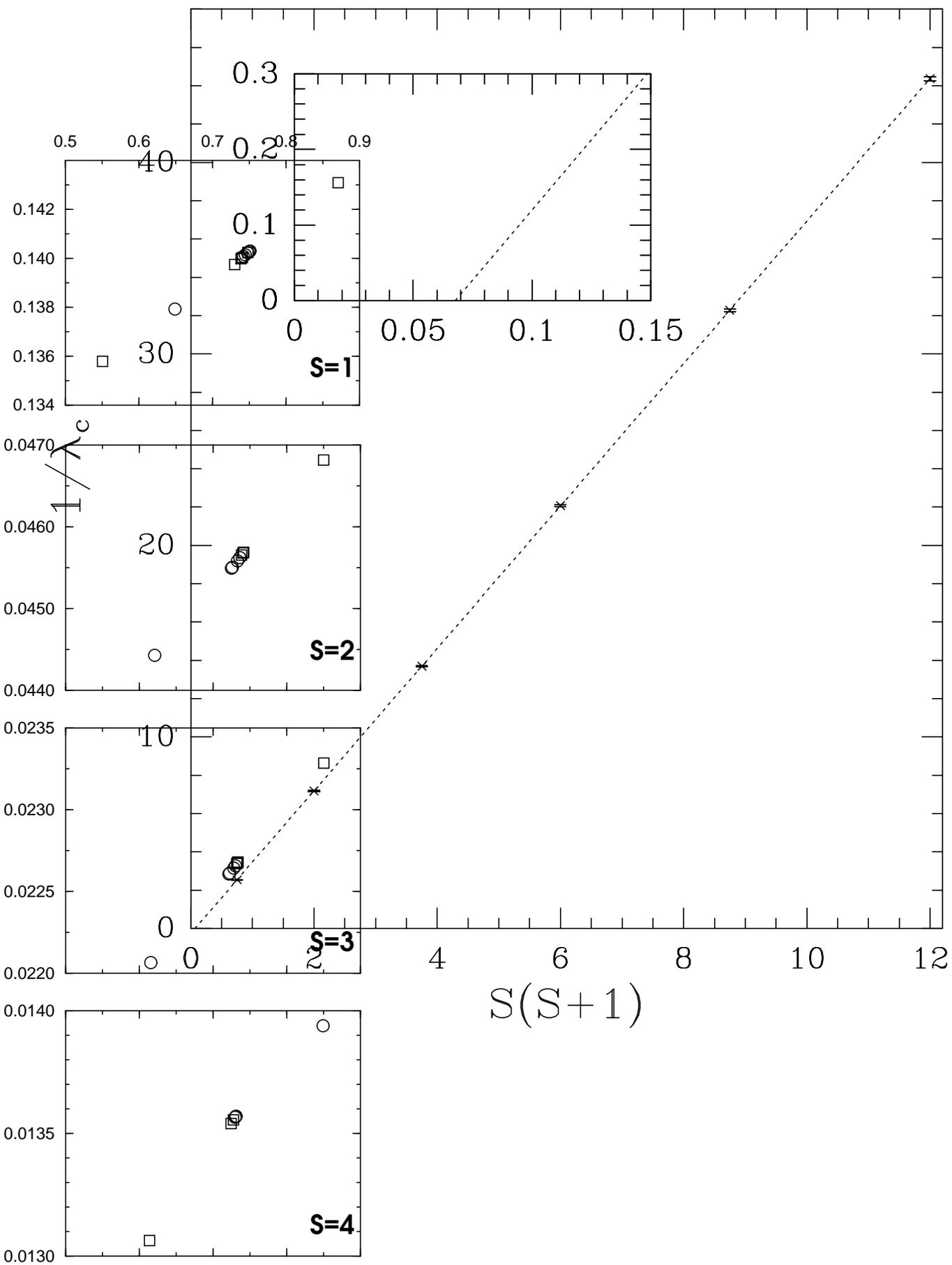
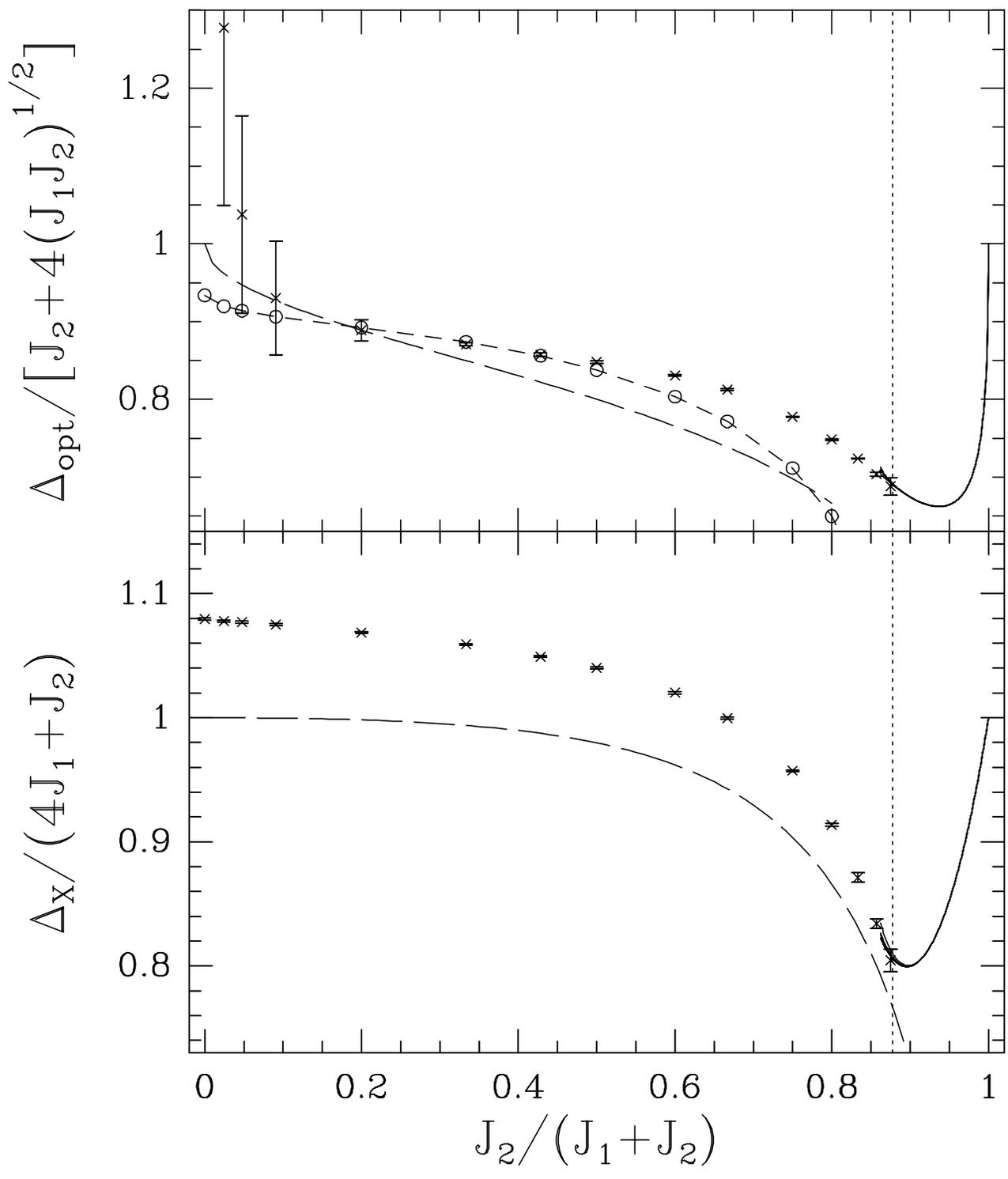
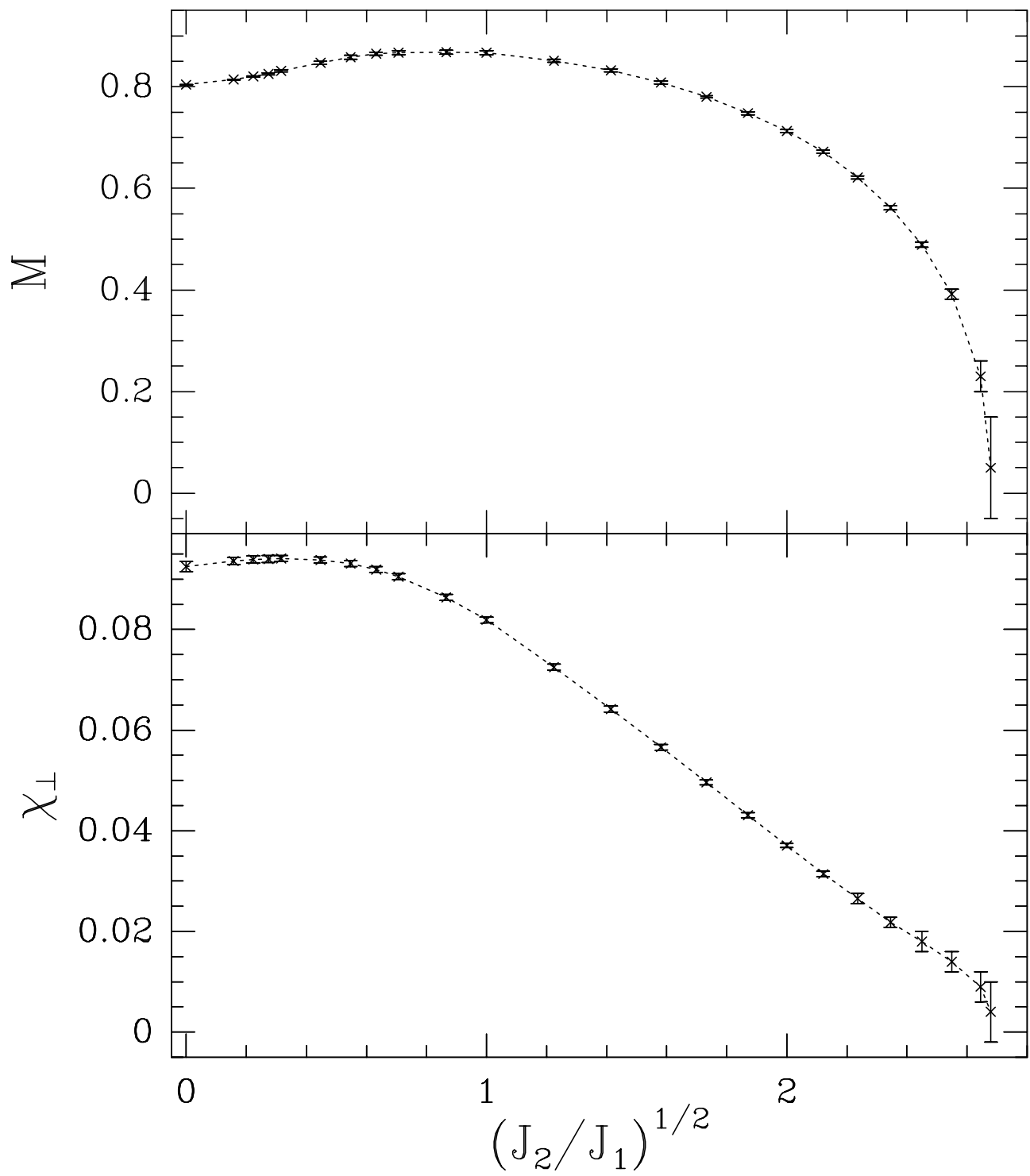
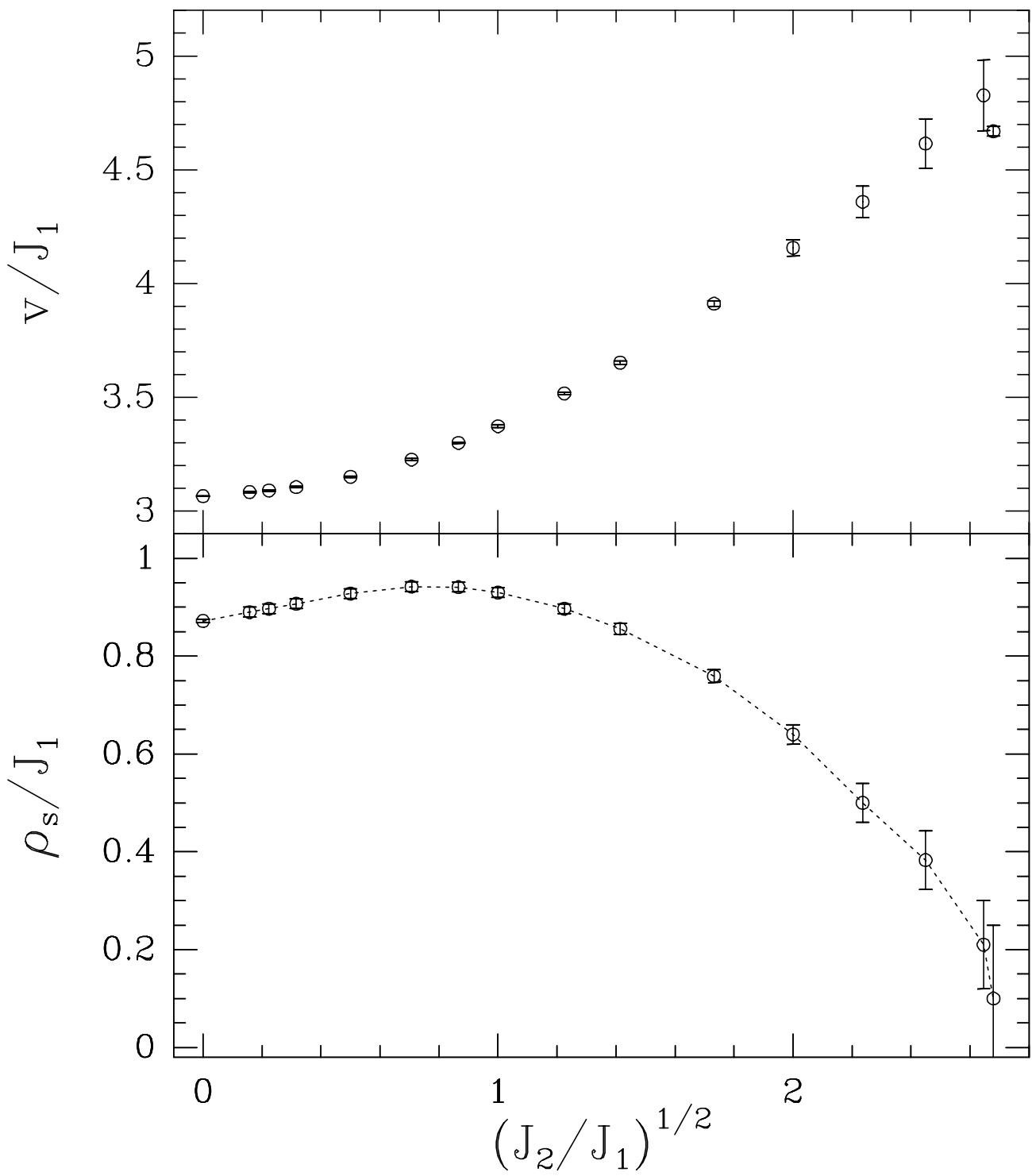
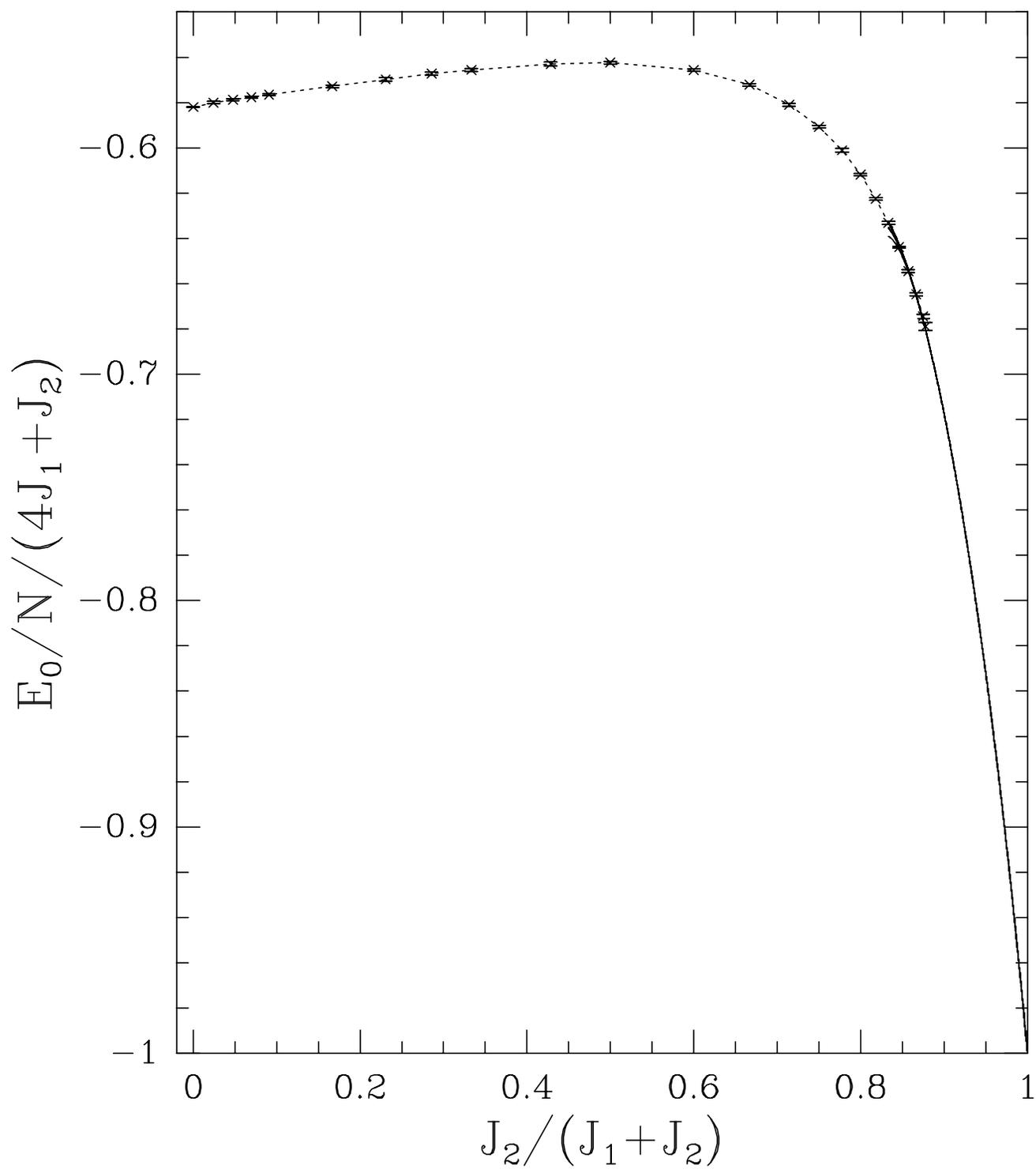
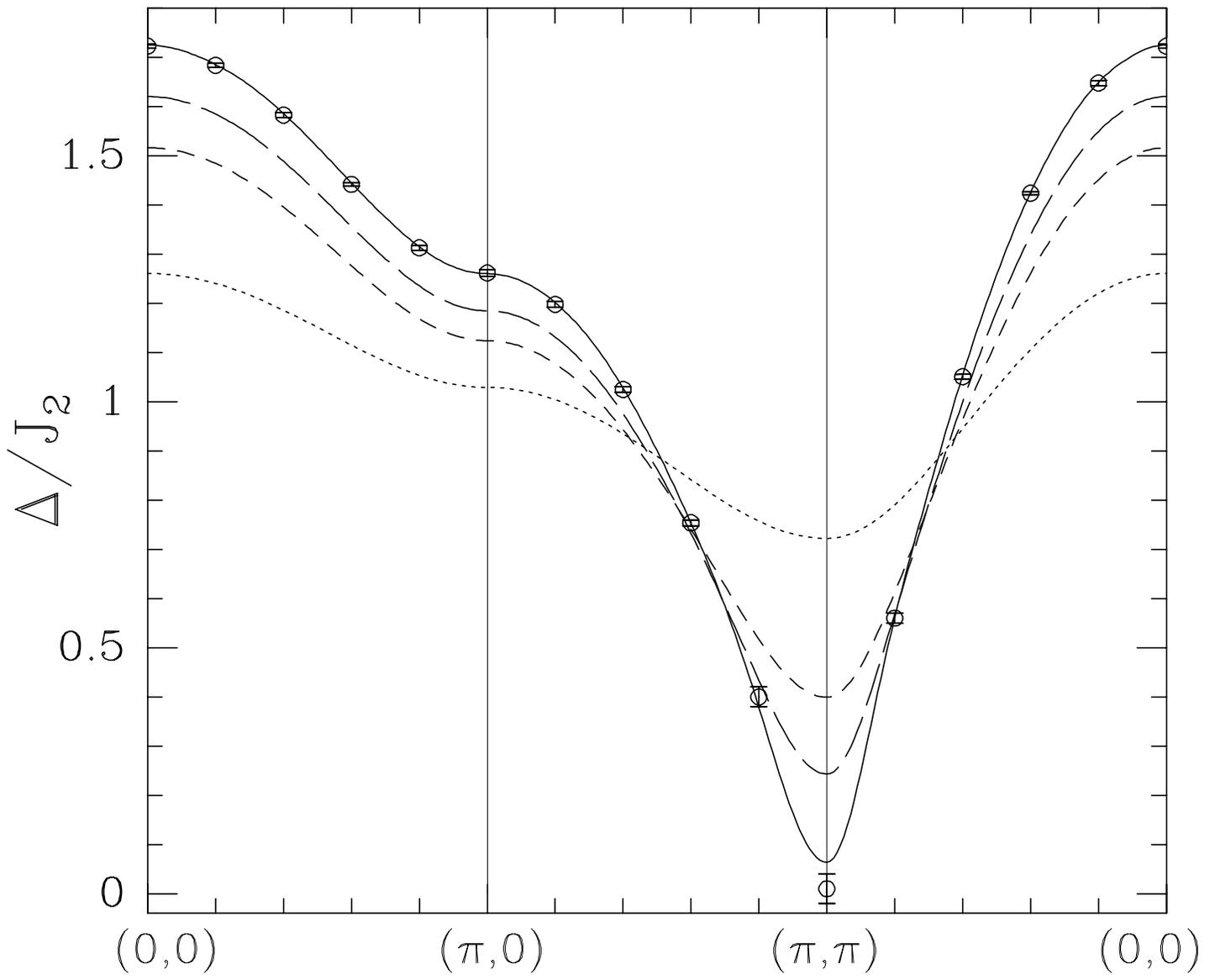
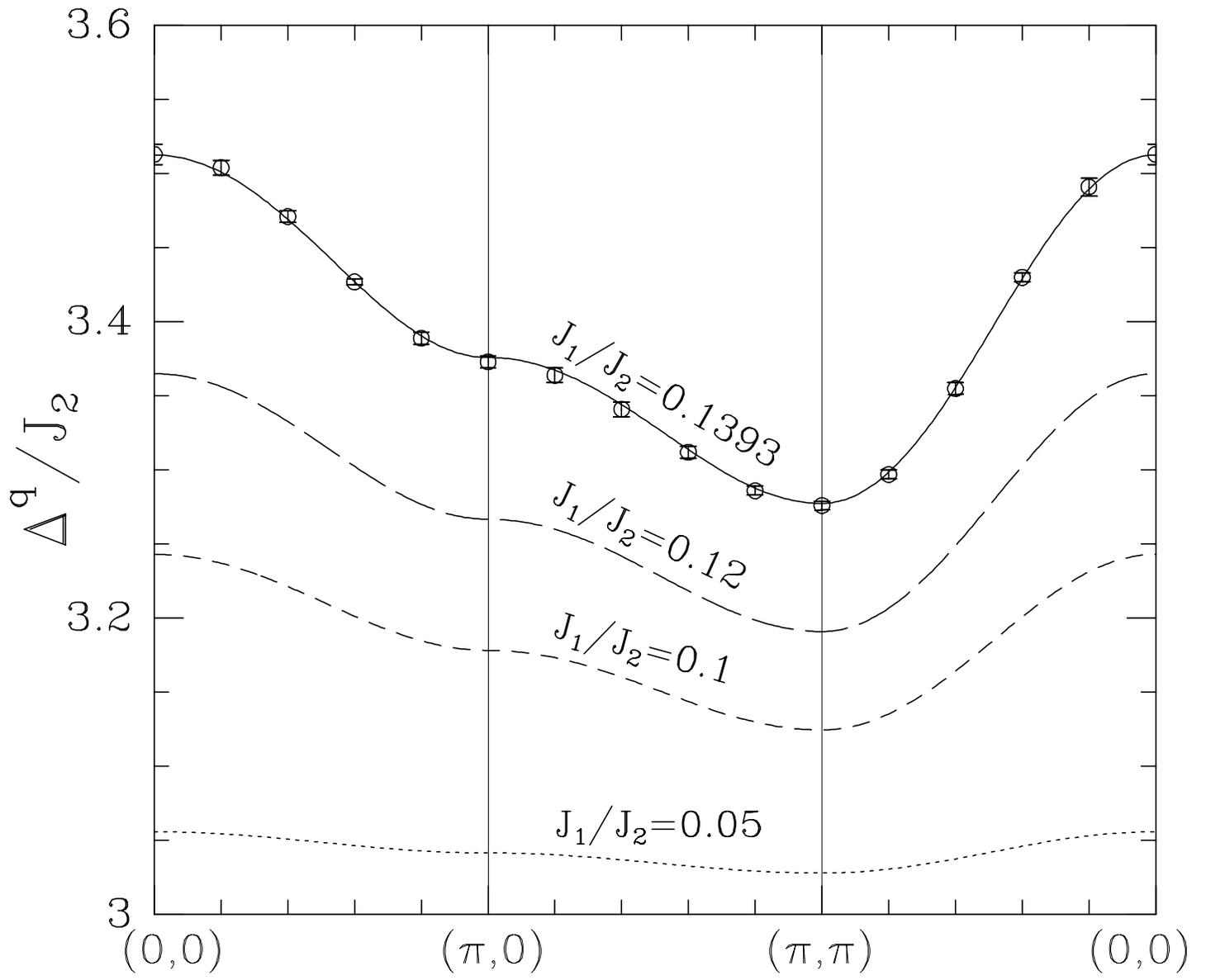
\end{document}